%% file: main.tex
\documentclass[11pt]{article}
\usepackage{graphicx}

\newcommand{\BABARPubYear}    {03}

\newcommand{\LANLNumber} {0000}

\input mymacro.tex

\input babarsym

\long\def\inst#1{\par\nobreak\kern 4pt\nobreak
    {\it #1}\par\vskip 10pt plus 3pt minus 3pt}

\begin{document}
{\pagestyle{empty}

\begin{flushleft}
% BAD NOTE 670, v5
\end{flushleft}
\begin{flushright}
\babar-CONF-\BABARPubYear/012 \\
% SLAC-PUB-\SLACPubNumber \\
hep-ex/\LANLNumber \\
July 2003 \\
\end{flushright}

\par\vskip 5cm

% Title of the paper
\begin{center}
\Large \bf Evidence for the Rare Decay $B\rightarrow \bf{J/\psi \eta K}$
\end{center}
\bigskip

\begin{center}
\large The \babar\ Collaboration\\
\mbox{ }\\
\today
\end{center}
\bigskip \bigskip

% Abstract
\begin{center}
\large \bf Abstract
\end{center}
\input abstract.tex
\vfill
\begin{center}
Presented at the 
International Europhysics Conference On High-Energy Physics (HEP 2003),
7/17---7/23/2003, Aachen, Germany
\end{center}

\vspace{1.0cm}
\begin{center}
{\em Stanford Linear Accelerator Center, Stanford University, 
Stanford, CA 94309} \\ \vspace{0.1cm}\hrule\vspace{0.1cm}
Work supported in part by Department of Energy contract DE-AC03-76SF00515.
\end{center}

\newpage
} % end of pagestyle{empty}

% Input author list file
%\input pubboard/authors_eps2003.tex
\input authors_eps2003.tex

% The body of the paper starts here

\input base.tex

\end{document}

%% file: mymacro.tex
\def\bjggkp{ \mathrm{B}^\pm \rightarrow \mathrm{J}/\psi \eta(\gamma\gamma) \mathrm{K}^\pm}

\def\bjggks{ \mathrm{B}^0 \rightarrow \mathrm{J}/\psi \eta(\gamma\gamma) \mathrm{K}^0_\mathrm{S}}

%% file: abstract.tex
We report evidence for the $B$ meson decays, 
$B^\pm\rightarrow J/\psi \eta K^\pm$ 
and 
$B^0\rightarrow J/\psi \eta K^0_S$, 
using {90} million $B\overline B$ events collected at the 
\FourS\ resonance with the \babar\ detector at the PEP-II $e^+ e^-$ asymmetric-energy storage ring. 
We obtain preliminary branching fractions 
in the charged and neutral channels of
$(10.8\pm 2.3(stat.)\pm 2.4(syst.))\times 10^{-5}$   and 
$(8.4\pm 2.6(stat.)\pm 2.7(syst.))\times 10^{-5}$, respectively. 

%% file: authors_eps2003.tex
\begin{center}
\small

The \babar\ Collaboration,
\bigskip

%% author list as of 02-Jun-2003 (595 authors)
%
B.~Aubert,
R.~Barate,
D.~Boutigny,
J.-M.~Gaillard,
A.~Hicheur,
Y.~Karyotakis,
J.~P.~Lees,
P.~Robbe,
V.~Tisserand,
A.~Zghiche
\inst{Laboratoire de Physique des Particules, F-74941 Annecy-le-Vieux, France }
% ...

\end{center}\newpage

%% file: base.tex
\begin{figure}[h]
%\centerline{\epsfxsize=4.0truein\epsffilprl-diagram.eps}}
\centerline{\includegraphics[height=7cm]{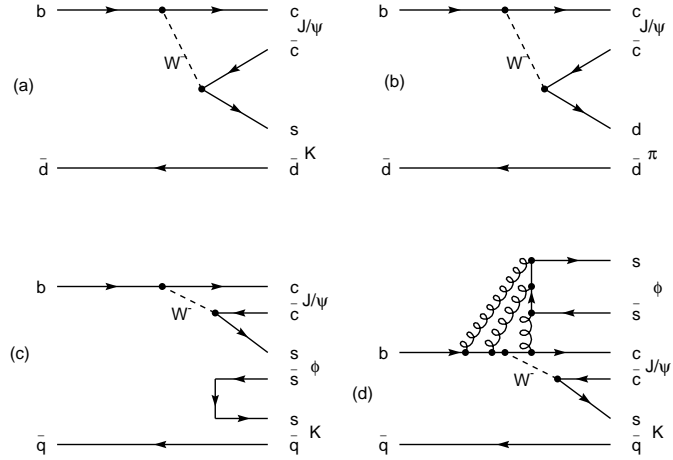}}
\vspace{0.0truein} \caption{The Feynman diagrams of 
color-suppressed B decays} \label{fig:feynman} \end{figure}

The first observation of color-suppressed B decay modes with hidden
strangeness, $s\overline{s}$, was in  the
decay mode, $B\rightarrow J/\psi \phi K$, by the CLEO
collaboration~\cite{ref;cleo} and more recently from the BaBar
collaboration~\cite{jpsiphik}. 
The color suppressed B decay
modes are represented at the parton level in
Figure~\ref{fig:feynman}.

\bigskip

Figures~\ref{fig:feynman}(a) and (b) show Feynman diagrams for
two-body final state color-suppressed B decays
into Cabbibo allowed ($J/\psi K$) and Cabbibo suppressed ($J/\psi \pi$) modes.
The color suppression is due to the requirement that
the color of the quark and anti-quark from the $W$ decay
must appropriately be matched with the charm quark and
the anti-d quark such that the final state particles
are color singlets.
Producing a
$\phi$ meson in a three-body final state may be achieved  by the
addition of $s\overline{s}$ quark-antiquark pairs as shown in
Figure~\ref{fig:feynman}~(c). It is possible also that a $\phi$ is
formed from three gluon emission as shown in
the OZI violating decay in
Figure~\ref{fig:feynman}(d). Since the $\eta$ meson is described
by the quark-antiquark combination,
\[
\eta =\frac{u\overline{u}+d\overline{d}-2s\overline{s}}{\sqrt{2}}
\]%
we would also expect B mesons, via the similar diagrams, to decay
into states with an $\eta $, such as $B\rightarrow J/\psi \eta K$, where the $s%
\overline{s}$ pairs or two gluons form an $\eta $ instead of a $\phi $.
Experimentally, the $\eta$ is seen in charmless decays at a
somewhat anomalously high rate~\cite{ref:eta_anomalous}, which
suggests that gluonic $\eta $ couplings are large. 
This may lead to
an enhancement in the decay rate of
$B\rightarrow J/\psi \eta K$. Since the basic diagrams are the
same, we would expect the branching fraction of this mode to be
comparable to BR($B\rightarrow J/\psi \phi K$)= $(4.4\pm
1.4\pm 0.5)\times 10^{-5}$~\cite{jpsiphik}. If the rate 
for non-resonant $B\rightarrow J/\psi \eta K$ is 
found to be much higher or
lower, then some other physics may be indicated.

If the mode $B\rightarrow J/\psi \eta K$ is observed, 
the neutral mode may be useful for
\CP\ violation studies~\cite{ckm},~\cite{cp-papers} 
and
investigation of quasi-two body decays.
The decay,
$B\rightarrow \psi (2s)K$, 
has been observed, and hence the
decay
$B\rightarrow \psi (2s)K\rightarrow J/\psi \eta K$
should be produced where the $J/\psi \eta$ are in
a relative P-wave state with
an inferred branching fraction~\cite{pdg} of 
$B(B^+\rightarrow \psi(2s)K^+\rightarrow J/\psi \eta K^+ )=(2.1\pm 0.2)\times 10^{-5}$.
If we observe
a two body S-wave J/$\psi \eta $ resonance, it would have the
quantum numbers, $J^{PC}=1^{+-}$, which are those of the
unconfirmed
$h_{c}$. However, since the mass threshold of J/$\psi
\eta $ is 3.64 GeV$/c^2$, which is higher than the 
lowest lying 
$h_{c}$ mass (estimated as the average of the three $\chi$ state
masses), such an observed resonance could indicate a new 
particle state. We note in addition that 
if a $J/\psi \eta$ resonance is found with a mass higher
than open charm threshold ($\sim$3.77 GeV/$c^2$)
it is unlikely to be a conventional charmonium state. 

The data used in this analysis 
correspond to a total integrated luminosity of 
$81.87$ fb$^{-1}$ taken on the $\FourS$
resonance producing a sample 
of $89.96\pm0.99$ million $B\overline{B}$ events ($N_{B
\overline{B}}$).
The data
were collected at
the PEP-II asymmetric-energy $e^{+}e^{-}$ storage ring with the $\babar$ detector, 
fully described elsewhere~\cite{babar-det}. The $\babar$
detector includes a five-layer silicon vertex tracker (SVT) and a 
forty-layer drift chamber (DCH) in a 1.5-T solenoidal magnetic field. 
These devices detect charged particles and measure their momentum and
energy loss. Photons and neutral hadrons are detected in a
CsI(Tl) crystal electromagnetic calorimeter (EMC). The EMC detects
photons with energies as low as 20 MeV and identifies electrons by their energy deposition. 
An internally reflecting ring-imaging
Cherenkov detector (DIRC), composed of quartz bars, 
measures the charged particle velocity for
particle
identification. 
Penetrating
muons and neutral hadrons are identified by the steel flux return, 
which is instrumented with 18-19 layers of resistive plate chambers (IFR).

The preliminary
selection criteria in this analysis follow previous
$\babar$ analyses~\cite{babar-charmonium}
and are briefly recalled here.
All charged track candidates are required to have at least 
12 DCH hits and transverse momentum greater than 100 MeV/$c$. 
The track candidates not associated with a $K^0_{S}$ decay must also
originate near the nominal beam spot. The muon, electron, and kaon candidates must have a polar angle in
radians of $0.3<\theta _{\mu }<2.7$, $0.410<\theta _{e}<2.409$, and $
0.45<\theta _{K}<2.50$, respectively.
In addition, all charged kaon candidates are required
to have a laboratory momentum greater than 250 MeV/$c$.
These requirements ensure the selection of tracks in the regions 
where the acceptance is well understood by the particle
identification (PID) systems.

The detailed explanation the 
PID is given elsewhere~\cite{e-mu-gam-pid},~\cite{kaon-pid}. 
Briefly,
photon candidates are identified from energy deposited 
in contiguous EMC crystals, summed together to form a cluster with total energy 
greater than 30 MeV and a shower shape consistent with that expected for 
electromagnetic showers. 
Electron candidates are required to have a good match 
between the expected and measured energy loss ($\dedx$) in the SVT and DCH, and 
between the expected and measured Cherenkov angle in the DIRC. 
The measurements of 
the ratio of EMC
shower energy to track momentum, and the number of EMC crystals associated with the track
candidate must be appropriate for an electron. 
Muons are selected based on the energy deposited in the EMC, 
the number and distribution of hits in the IFR,  the match between the IFR hits 
and the extrapolation of the track by the DCH 
into the IFR, and the depth of penetration of the track into the IFR.
Charged kaon candidates are selected based on energy loss
information from the SVT and DCH and the Cherenkov angle measured by the DIRC.

The intermediate states in the decay modes used in this analysis, 
$J/\psi \left(e^+ e^- \right) $, 
$J/\psi \left(\mu^+  \mu^- \right) $, 
 and $K_{S}^{0}\left( \pi ^{+}\pi ^{-}\right) $, are selected within the mass 
intervals listed in Table~\ref{table-mass}.
The di-electron mass interval is larger than the di-muon to 
account for Bremstrahlung in the detector.
The $K_{S}^0$ decay length is required to be greater than 0.1 cm. 

The remaining four selection criteria are 
the $\eta$ mass,
the $\pi^0$ veto, the photon helicity angle~\cite{babar-charmonium} from
the $\eta$ decay and the thrust angle.
In the following
we describe each of these criteria and the
process used to determine 
their final cut values.

The $\eta$ candidates are required to have $\gamma\gamma$
mass within the range listed in Table~\ref{table-mass}.
If either of the photons associated with an $\eta$ candidate,
in combination with any other photon in the event,
forms
a $\gamma \gamma $ mass  near the nominal $\pi^0$ mass, 
the $\eta$ candidate is vetoed as a $\pi^0$ background.
The $\pi^0$ veto selection was adjusted by limiting the
mass difference between all possible $\gamma \gamma$
masses and the nominal $\pi^0$ mass as shown Table~\ref{table-mass}. 
The $\eta$ candidate is rejected if 
$\left| \cos \theta _{\gamma }^{\eta }\right|$ is near 1, where $
\theta _{\gamma }^{\eta }$ is the photon helicity angle in the $\eta$ rest
frame. 
This rejects combinatoric background due to random pairs of photons
that typically have a  photon helicity angle distribution that
peaks at 0 or 180 degrees. 

An additional requirement is applied to separate two-jet continuum events from the
more spherical $B$ meson decays. The angle $\theta _{T}$ between the 
thrust~\cite{babar-charmonium} direction of the $B$
meson candidate and the thrust direction of the remaining tracks in the
event is calculated. We reject events
when  $\left| \cos \theta _{T}\right|$ is near 1, 
since 
%these thrust axes are uncorrelated and
the distribution 
in $\cos\theta _{T}$ 
is flat for $B\overline{B}$ events, 
while 
for background continuum events 
the distribution is peaked at $\cos \theta _{T}=\pm 1$.

\begin{figure}[!htb]
\begin{center}
\includegraphics[height=7cm]{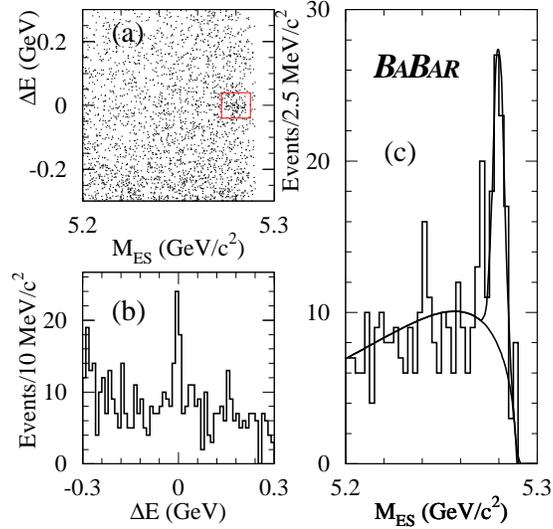}
\caption{The $\Delta E$ and $\mes$ distributions for $B^{+}\rightarrow J/\psi \eta K^+$. 
The $\Delta E$ vs. $\mes$ event distribution is shown in (a) with a 
small rectangle corresponding to the signal region selection defined
in the text. 
The $\Delta E$ projection within a $\mes$ signal region selection is shown in (b).
The $\mes$ projection within a $\Delta E$ signal region selection is shown in (c).
The solid line in (c) is the fit
described in the text.} 
\label{fig:jpkub}
\end{center}
\end{figure}

\begin{figure}[!htb]
\begin{center}
\includegraphics[height=7cm]{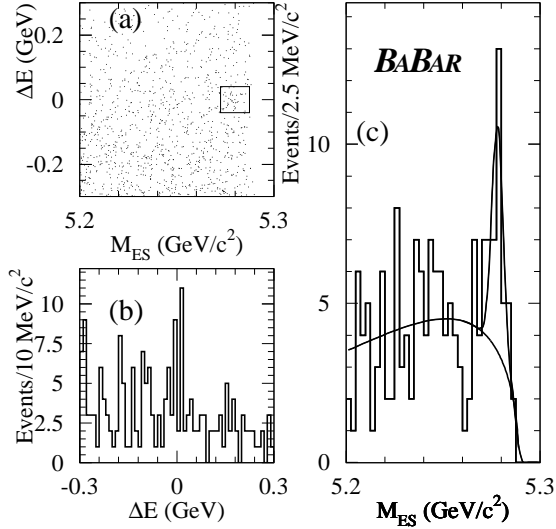}
\caption{The $\Delta E$ and $\mes$ distributions for $B^{0}\rightarrow J/\psi \eta K^0_S$. 
The descriptions of Figs. 3(a), (b) and (c) follow those of Figs. 2(a), (b) and (c), respectively.}
\label{fig:jpksub}
\end{center}
\end{figure}

Estimation of the signal and the background use two kinematic variables: 
the energy difference $\Delta E$ between the energy of the $B$ candidate and 
the beam energy $E_{b}^{*}$ in the $\FourS$ rest frame; and the energy-substituted 
mass $\mes =\sqrt{\left( E_{b}^{*}\right)^2 -\left( P_{B}^{*}\right) ^{2}}$, where 
$ P_{B}^{*}$ is the reconstructed momentum of the $B$ candidate in the $\FourS$ frame.
Typically these two variables in a two-dimensional plot
for the $B$ meson signal will appear as a
weakly correlated two-dimensional Gaussian distribution,
whereas the background is roughly uniformly distributed.
The $\Delta E$ and $m_{ES}$ resolutions are mode dependent.
A signal region for each mode, $B^+$ and $B^0$, is defined as a rectangular region in the 
$\Delta E$ versus $\mes$ plane with
$|\mes-m_{B}|<7.5$ MeV/c$^2$, where $m_{B}$ is the mass of $B$ meson
and $|\Delta E|<$40 MeV. 
Before the data were analyzed, these four final selection criteria were optimized using 
a Monte Carlo (MC) simulation of the signal and the known backgrounds.
Motivated by the $B\rightarrow J/\psi \phi K$ measurement, the
$ab\ initio$ value of the
branching fraction for $B\rightarrow J/\psi \eta K$ 
used in the signal MC was $5\times 10^{-5}$.
The number of reconstructed MC signal events ($n_s^{mc}$) and the number of
reconstructed MC background events ($n_b^{mc}$) in the signal box were obtained using the same 
criteria to
estimate the signal significance ratio, $n_s^{mc}/ \sqrt{n_s^{mc}+n_b^{mc}}$. 
This ratio was maximized by
varying these four selection criteria.
The resulting criteria were fixed and are listed in Table~\ref{tab:optcuts}.
The resulting $\Delta E$ and $\mes$ distributions for data 
are shown in 
Figs.~\ref{fig:jpkub} and \ref{fig:jpksub}. 
The number of data events ($n_0$) observed in the
signal box region for each mode, $B^+$ and $B^0$, is listed in Table~\ref{table-upperlimit-bf}. 

%%%%%%%%%%%%%%%%%%%%%%%%%%%%%%%%%%%%%%%%%%%%%%%%% figures

The 
$\mes$ distribution background shape is determined by fitting 
the line shape of an
ARGUS function~\cite{babar-charmonium}
for each mode   
to the $\mes$ distributions  
formed from the $\Delta E$ sideband region of the on-peak data 
of $0.1<|\DeltaE|<0.14$ for the $B^+$ mode
and 
$0.08<|\DeltaE|<0.28$ for the $B^0$ mode.
To determine the number of signal events, we fit the data $\mes$ distribution 
to a model
( see Figs.~\ref{fig:jpkub}(c)  and \ref{fig:jpksub}(c) ) 
that is the sum of
this ARGUS function
whose normalization is allowed to vary,
and 
a signal Gaussian, whose width is fixed to the MC value
but whose mean and normalization are allowed to vary.
The integral of this resulting ARGUS function over the signal box region 
yields 
the number of background events 
($n_{b}$)
and its uncertainty ($\sigma_b$).
Similar $\mes$ background shapes were found by using a $\gamma\gamma$ mass
sideband outside the nominal $\eta$ mass. 
Occasionally backgrounds with an inclusive $J/\psi$ decay will
lead to a peaking background near or at the $B$ mass in
the $m_{ES}$ distribution. 
Background shapes consisting of a sum of the ARGUS function and a broad
Gaussian were attempted, however its contribution was
found to be small and was dropped from the final background fit.
The final values of $n_{b}$ and $\sigma_b$ are listed in Table~\ref{table-upperlimit-bf}.

We find evidence for signals in the 
$J/\psi \eta K^{+}$ and $J/\psi \eta K_{S}^{0}$ modes
as seen in  
Figs.~\ref{fig:jpkub} and \ref{fig:jpksub}. 
To determine the
branching fraction for these modes 
the number of signal events ($n_s$) is given by a simple subtraction 
of the estimated number of background events
from the events in the signal box region, 
$n_{s}=n_{0}-n_b$. 
The efficiencies ($\epsilon$) for each mode, listed in 
Table~\ref{table-upperlimit-bf},
are determined by MC simulation 
with three--body phase space and
unpolarized $J/\psi$ decays.
The calculation of the branching fraction ($BF$)
is 
\begin{eqnarray*}
BF &=&\frac{n_\mathrm{s}}{N_{B\overline{B}}\times \epsilon \times f} \\
\end{eqnarray*}
where $\epsilon$ is the efficiency for the mode and $f$ is the
product of secondary branching fractions
for the $J/\psi ,$ $\eta $, and $K^0_S$.
The results are given 
in the last column of Table~\ref{table-upperlimit-bf} where 
the first and second $BF$ uncertainities are the statistical and
systematic, respectively.
The statistical uncertainty is derived from the error
on $n_s$ which is $\sqrt{n_0+\sigma_b^2}$.

In Table~\ref{table-systematics}, we list the contribution to the systematic  
uncertainty
from the error 
on each of the following quantities: 
$N_{B\overline B}$; 
secondary branching fractions~\cite{pdg}; 
MC statistics; 
PID, tracking, and photon detection efficiencies; 
$\pi^0$ veto;
$\eta$ mass range;
background parameterization;
and
model dependence.
The PID, tracking, and photon detection efficiency 
uncertainties are based on the study of data control 
samples~\cite{con-sam}.
The error in the $\pi^0$ veto efficiency
was studied by measuring the inclusive $\eta$ rate
in both data and MC. After varying the
$\pi^0$ veto mass cut, the change in 
the number of inclusive $\eta$'s was determined
in data and MC to measure the deviation between
the data and the simulation and 
to estimate the resulting systematic uncertainty.
The uncertainty in the $\eta$ mass range was determined by
comparing the measured $\eta$ mass resolution in inclusive $\eta$
decays to the $\eta$ mass resolution from the signal MC.
The background parameterization uncertainty was
estimated by changing the ARGUS shape parameter
by $\pm$1 standard deviation, refitting the $\mes$ data distribution
and recalculating the number of signal events. 
Additional systematic uncertainties due to the decay model dependence are 
estimated for the modes $J/\psi \eta K^{+}$ and $J/\psi \eta K_{S}^{0}$. 
MC simulations are used to determine how much  the efficiency depends 
on assumptions about intermediate resonances and angular distributions. 
Five samples were studied.
One sample is  generated with $100\%$ transversely polarized $J/\psi$  
and another with $100\%$ longitudinally polarized $J/\psi$. 
The other three samples had large $J/\psi \eta$ mass, large $\eta K$ mass
or small $J/\psi K$ mass.
The resulting relative change in efficiency is entered as a fractional systematic 
uncertainty in Table~\ref{table-systematics}. 
The total systematic uncertainty for each mode combines all these 
separate errors in quadrature,
is listed the last column in Table~\ref{table-systematics}. 
and is used to determine
the $BF$ systematic uncertainties listed in
Table~\ref{table-upperlimit-bf}.

The probability for a null hypothesis (P-value) 
is defined as
the Poisson probability that the estimated number of background 
events fluctuates to observed number of events $n_0$ or greater. 
In
Table~\ref{table-upperlimit-bf} we provide
P-values 
calculated
using the central value of the background estimate $n_b$ and the value increased by 
1 standard deviation, 
$n_b + \sigma_b$, to provide an estimate of probability including the background systematic uncertainty. 
%
%
%Upper limits are set as well as branching fractions are calculated 
%for the modes with no signal or limited statistical evidence 

We determine also the 90$\%$ confidence level upper limit 
on the branching fraction using
 $n_{0}$, $n_{b}$, and $\sigma_b$, 
in the signal region, and the total systematic uncertainty $\sigma _{T}$. 
Assuming the two uncertainties ($\sigma _{b},\sigma _{T}$) are uncorrelated and Gaussian, 
the Bayesian upper limit on the number of events ($N_{90\%}$) is obtained by
folding the Poisson distribution with two normal distributions describing  
these two uncertainties and integrating the resulting function 
to the $90\%$ confidence level (C.L.). 
This assumes that the $a\ priori$  branching fraction distributions are uniform.
The results are listed in
Table~\ref{table-upperlimit-bf}.

Our preliminary branching fraction is comparable to 
$B\rightarrow J/\psi \phi K$ and is consistent 
with simple expectations from color-suppressed decay diagrams.
The ratio of the charged $(J/\psi \eta K^\pm)$
to neutral $(J/\psi \eta K_{S}^{0})$
branching fractions is
expected to be two which is not
inconsistent with our results.
Although the current statistics are too small,
a larger data set of the neutral mode could be useful
for a test of $CP$ violation 
and both charged and neutral modes may enable
a search for resonances in the
two body intermediate states.

In summary, we find evidence
for the decay 
$B\rightarrow$ $J/\psi \eta K$ in two
modes with preliminary
branching fractions of, ${\cal B}$($B^+\rightarrow J/\psi \eta K^{+}$) =
{${(10.8\pm 2.3\pm 2.4)\times 10}^{-5}$} and 
${\cal B}$($B^0\rightarrow J/\psi
\eta K_{S}^{0}$) = {$(8.4\pm 2.6\pm 2.7)\times 10^{-5}$}. 

\label{sec:Acknowledgments}
\input acknow_PRL.tex

%%%%%%%%%%%%%%%%%%%%%%%%%%%%%%%%%%%%%%%%%%% start tables

\pagebreak
\begin{center}
\Large {\bf TABLES} 
\end{center}

\begin{table*}[!htb]
\caption{Mass regions for selection of intermediate particles.}
\begin{center}
\begin{tabular}{lrcccl}
\hline\hline
Mode & \multicolumn{5}{c}{Mass Range (GeV/$c^{2}$)} \\ \hline
$J/\psi \rightarrow e^+e^-$ & $2.95$&$<$&$M(e^{+}e^{-})$&$<$&$3.14$ \\ 
$J/\psi \rightarrow \mu^+ \mu^- $ & $3.06$&$<$&$M\left( \mu ^{+}\mu ^{-}\right)$&$ <$&$3.14$ \\ 
$K_{S}^{0}\rightarrow \pi ^{+}\pi ^{-}$ & $0.489$&$<$&$M\left( \pi ^{+}\pi^{-}\right)$&$ <$&$0.507$ \\ 
$\eta \rightarrow \gamma \gamma $ & $0.525$&$<$&$M\left( \gamma \gamma \right) $&$<$&$0.571 $ \\ \hline
\end{tabular}
\end{center}
\label{table-mass}
\end{table*}

\begin{table*}[htb]
\begin{center}
\caption{Final selection criteria for the $\bjggkp$ and $\bjggks$ modes.}
\label{tab:optcuts}
\begin{tabular}{lcc}
\hline\hline\\[-0.2cm]
%\hline
Variable &$J/\psi \eta K^+$&$J/\psi \eta K^0_S$\\\hline
$ | M_{\eta} - 0.547 | \leq $ & 0.023 GeV/c$^2$ & 0.023
GeV/c$^2$\\
$\pi^0$ veto if $ | M(\gamma_\eta+\gamma_{other}) - 0.135 | \leq $ & 0.017
GeV/c$^2$& 0.010 GeV/c$^2$\\
Helicity: $ | cos(\theta_\gamma^\eta) | \leq $& 0.93 & 0.81\\
Thrust: $ | cos(\theta_T) | \leq $& 0.8& 0.9\\ \hline
\end{tabular}
\end{center}
\end{table*}

\begin{table*} [!htb]
\caption{Branching fractions and 90\% C.L. upper limits.}
\begin{center}
\footnotesize{
\begin{tabular}{lccccccc}
\hline\hline\\[-0.2cm]

Mode & $\epsilon$ & $n_{0}$ & $n_{b}\pm \sigma _{b}$ & $N_{90\%}$& $90\%$ C.L.U.L.& P-value  & ~~~Branching Fraction \\ 
& &&&& $(10^{-5})$ & range & $(10^{-5})$\\\hline \\[-0.2cm]

$J/\psi \eta K^{+}$ &  $10.75\%$&99 & $50.3\pm 3.0$ &70.0&$<$15.5&$(0.09-1.42)\times 10^{-8}$&
$10.81\pm 2.31\pm 2.37$ \\ 
$J/\psi \eta K_{S}^{0}$ & ${8.53\%}$&39 & {${18.5\pm 1.7}$} &34.5&$<$14.1&$(.23-1.3)\times 10^{-4}$& 
~~~$8.35\pm 2.64\pm 2.67$ \\ \hline 
\end{tabular}
}
\end{center}
\label{table-upperlimit-bf}
\end{table*}

\begin{table*}[!htb]
\caption{Systematic uncertainty summary on the branching fractions. All are fractional uncertainties in percent.}
\begin{center}
\footnotesize{
\begin{tabular}{lccccccccc}
\hline\hline\\[-0.2cm]
Mode &$N_{B\overline B}$ & Secondary & M.C. & PID, &$\pi ^0$& $\eta$ & Bkgd.&Model & Total \\
& & Branching &Stat.& tracking, & veto & mass & Param.&& \\
& & Fraction&    &$\gamma$ Det.&  & cut& && \\\hline\\[-0.2cm]
$J/\psi \eta K^+$ &1.1& 2.48 & 1.77 & 8.2 & 8.1& 3.40& 16.7 &5.1& 22.0 \\ 
$J/\psi \eta K_{S}^0$ &1.1& 2.52 & 2.17 & 8.3 & 8.3 & 3.14&27.0&9.5 & 32.0 \\ \hline
\end{tabular}
}
\end{center}
\label{table-systematics}
\end{table*}

%%%%%%%%%%%%%%%%%%%%%%%%%%%%%%%%%%%%%%%%%%%%%%%%%%%%%%%%%%%

%% file: acknow_PRL.tex
We are grateful for the excellent luminosity and machine conditions
provided by our \pep2\ colleagues, 
and for the substantial dedicated effort from
the computing organizations that support \babar.
The collaborating institutions wish to thank 
SLAC for its support and kind hospitality. 
This work is supported by
DOE
and NSF (USA),
NSERC (Canada),
IHEP (China),
CEA and
CNRS-IN2P3
(France),
BMBF and DFG
(Germany),
INFN (Italy),
FOM (The Netherlands),
NFR (Norway),
MIST (Russia), and
PPARC (United Kingdom). 
Individuals have received support from the 
A.~P.~Sloan Foundation, 
Research Corporation,
and Alexander von Humboldt Foundation.

%% file: main.bbl
\begin{thebibliography}{99}
\bibitem{e-mu-gam-pid} The particle identification selection
is taken from ref. ~\cite{babar-charmonium}. 
The photon candidates selection is the same 
as that in ref.~\cite{babar-charmonium}.
The electron candidates 
are required to satisfy
the ``Tight" selections and
the muon candidates use
the ``Loose" selections,
as specified in ref.~\cite{babar-charmonium}.
\bibitem{con-sam} The estimate of the uncertainty on 
the PID, tracking and photon detection
efficiency using control samples from data is
described in section XI of ref.~\cite{babar-charmonium}.
\end{thebibliography}
